\documentclass{nature}
\usepackage{cite}
\usepackage{color}
\usepackage{textgreek}
\usepackage{xcolor}
\usepackage[load-configurations = abbreviations]{siunitx}
\renewenvironment{figure}{\let\caption\NAT@figcaption}{}
\usepackage{graphicx}
\makeatletter
\let\saved@includegraphics\includegraphics
\AtBeginDocument{\let\includegraphics\saved@includegraphics}
\renewenvironment*{figure}{\@float{figure}}{\end@float}
\makeatother

\title{On droplets coalescence in a quasi-2D fluid}

\author{Christoph Klopp, and Alexey Eremin}

\begin{document}

\maketitle

\begin{affiliations}
 \item Otto von Guericke University Magdeburg, Institute of Physics, 39106 Magdeburg, Germany
\end{affiliations}

\begin{abstract}
The coalescence of droplets plays a crucial role in nature and modern technology. Various experimental and theoretical studies explored droplet dynamics in 3D and on 2D solid or liquid substrates. In this paper, we demonstrate the coalescence of isotropic droplets confined in thin quasi-2D liquids --  overheated smectic films. We observe the merging of micrometer-sized flat droplets using high-speed-imaging and analyse the shape transformations of the droplets on the timescale of milliseconds. Our studies reveal the scaling laws of the coalescence time, which exhibits a different dependence on the droplet geometry than in case of droplets on a solid substrate. A theoretical model is proposed to explain the difference in behaviour. 
\end{abstract}

\section*{Introduction}
Dynamics of floating objects and their interactions with the flow in restricted geometries is of paramount interest in physical, chemical and biological systems \cite{Saffman:2003ve, Mech:ft, Eremin:2011hk}. Furthermore, the coalescence and merging of liquid objects such as liquid droplets has a large practical relevance \cite{Wijshoff:2018cq,Ahmadlouydarab:2016bn,Narhe:2008co,Wasan:2000en,Mech:ft,Eggers:1999dz, {Watson:2014fe}}. One can find coalescence of droplets in everyday life such as during rain drop formation \cite{Low:1982ew}, it occurs in colloidal emulsions like mayonnaise and creams \cite{Goff:1997ij,SanjeevKumar:1996kj}, one  can also observe the merging of powders into a homogeneous material by heating (sintering). Industrial applications, such as ink-jet printing \cite{Ihnen:2012hb}, coating processes and the emulsions stability \cite{Eslamian:2018ft}, oil recovery etc. \cite{Perazzo:2018il},  require a better and detailed understanding of the merging dynamics of liquid droplets. 
The droplet coalescence is a complex phenomenon involving different dynamic regimes of the flow and the interactions with interfaces and the environment \cite{Mech:ft}. Pioneering studies of the droplet coalescence date back to the works of Lord Rayleigh in the nineteenth century \cite{Rayleigh:1879uv,Rayleigh:1879vw}. Many experimental and theoretical investigations of the 3D and 2D coalescence of liquid objects have been performed during the last two decades to fully understand the merging process of liquid objects \cite{Khodabocus:2018dw,Paulsen:2011cg,Sellier:2009kz,Sellier:2009fe,Thoroddsen:2007fs,Aarts:2005kt,Narhe:2004fn,Gaskell:2004ff,Eggers:1999dz,Hopper:1991cb,Mech:ft}. 

The coalescence of 3D droplets exhibits of three different dynamic regimes such as the inertial limited viscous regime at the very early stage of the neck formation, viscous regime dominating at the sufficiently early time of the neck formation, the inertial regime occurring at later times \cite{Paulsen:2012kq,Paulsen:2011cg, {Aarts:2005kt}}. Interactions with solid and viscoelastic substrate strongly affect the dynamics of the coalescing droplets \cite{Khodabocus:2018dw,Ahmadlouydarab:2016bn, {Eddi:2013ko},{Kapur:2007bo},{Schwartz:1998gw}}.

Self-similar dynamics was demonstrated by Hern\'andez-Sa\'nchez \cite{HernandezSanchez:2012fp} for the asymmetric droplet coalescence on a solid substrate. On the soft gel substrate the droplets exhibit long-range interactions, which is analogue to the Cheerios effect \cite{Karpitschka:2016fe}. In addition Hack et al. \cite{Hack:2020gj} recently investigated the coalescence of liquid droplets on a liquid surface and proposed a model to describe the behaviour and the scaling in the viscous and inertia regimes. The model was validated in materials with different viscosities exhibiting both coalescence regimes \cite{Hack:2020gj}. 

The theory of the coalescence in 2D (cylinders or discs) was proposed by Hopper \cite{Hopper:1991cb}. He developed an exact analytical solution of the problem. In contrast to the 3D geometry, the line tension drives the coalescence in 2D, which is accompanied by the planar flows. Experimental realisations of 2D coalescence in freely suspended liquid crystal films were made by Dolganov et al. \cite{Shuravin:2019eu} in air and  on a liquid water substrate by Delabre et al. \cite{Delabre:2010ce}.

In the present paper, we explore the coalescence dynamics of flat liquid droplets in a thin smectic-A film. Having the thickness of several molecular layers and the size of a few millimetres, such films are remarkable for their two-dimensional character of the flow, which makes them the examples of quasi-2D fluids \cite{{Schuring:2002bf},Eremin:2011hk,PhysRevFluids.2.124202,ISI:000404793400019}. At the same time, the coalescence of droplets is driven by the surface tension as in the 3D case. The micrometer-sized, lens-shaped droplets are embedded in a nanometer thin freely suspended smectic-A film \cite{Schuring:2002bf}. Employing the high-speed imaging with the interferometry techniques, we can monitor the coalescence dynamics on a sub-millisecond scale, and reconstruct the 3D shape changes of the droplets  in great detail.
In contrast to droplets on substrates or liquid surfaces, our droplets are not constrained by immobile surfaces. The material flow, can be considered as quasi two-dimensional. That allows us to describe the flow dynamics during the coalescence using the lubrication theory. 

\section*{Results}

\begin{figure}[h!]
	\centering
	\includegraphics[width=0.9\linewidth]{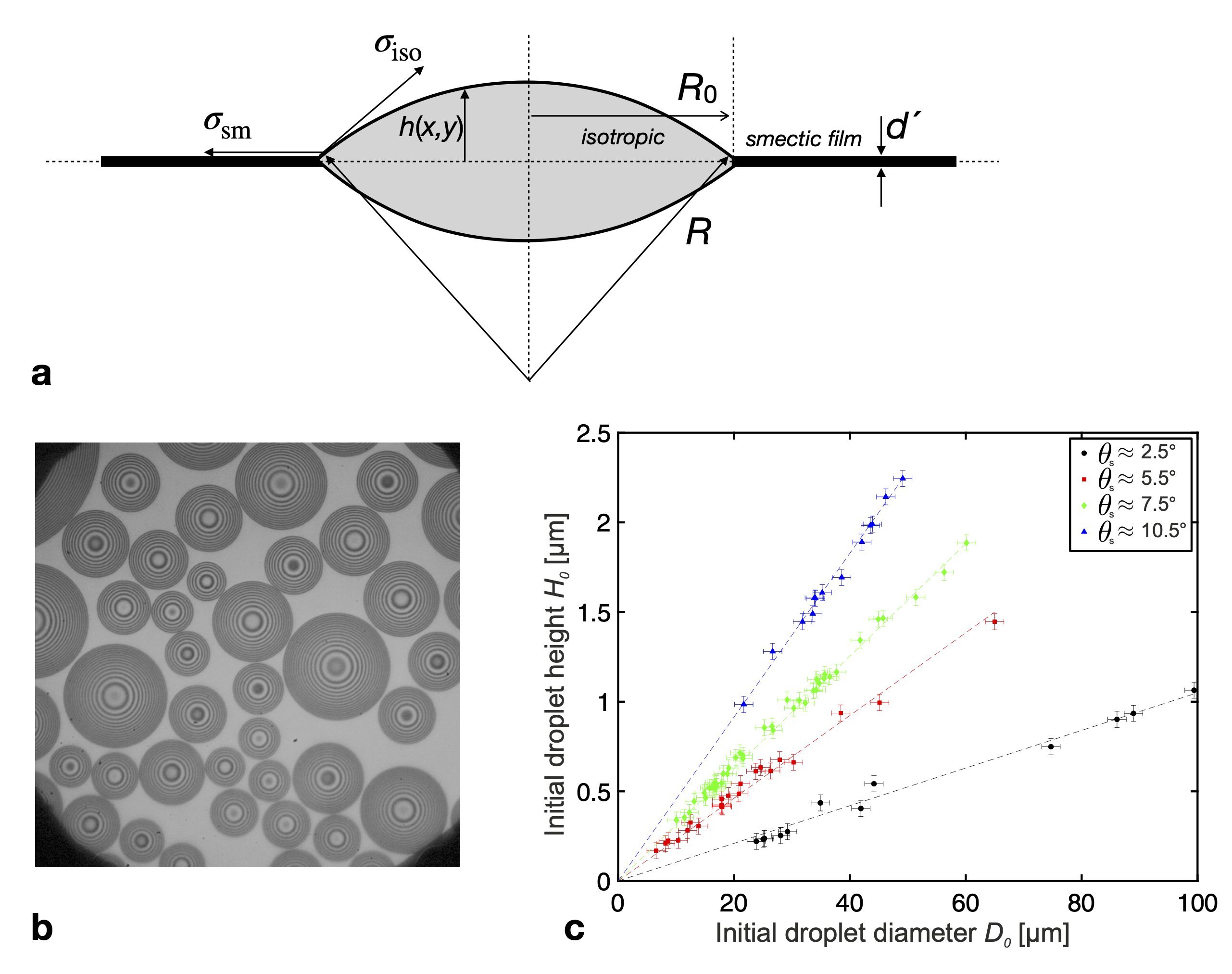}
	\caption{(a) A schematic of a droplet in a freely suspended film of thickness $d'$. The droplets have shapes of spherical cap pairs with large curvature radii of $R>>R_0$. The surface of the droplet is covered with a smectic film of a different thickness $d''$. (b) Isotropic droplets nucleated at a temperature $T=54^\circ C$ in a film of thickness roughly $d'\approx 1200$ nm observed in monochromatic light of wavelength $\lambda=546$ nm to visualise interference rings. (c) Dependence of the initial heights of droplets on the  initial diameter $D_0$ for different contact angles tuned by the temperature in the overheated films.}
	\label{fig:figure1}
\end{figure}

Isotropic droplets nucleate on heating above the clearing point of the liquid crystal in bulk (smectic-A to Iso) in freely suspended smectic-A films with thicknesses in a range of 50 -- 1500 nm. Reaching the clearing point, the melting material forms isotropic droplets with sizes varying from 5 --  100 \textmu m in diameter and 0.5 -- 2 \textmu m in height (Figures~\ref{fig:figure1}a and b). A thin smectic layer wets the droplet surface, which makes the surface tension slightly different from the tension of the freely suspended film. The droplets have lentil shapes determined by the difference between the interfacial tensions at the smectic/air and smectic/isotropic interfaces. 
The equilibrium shape of the freely floating droplets in FSFs is described in detail in \cite{Schuring:2002bf}. The height of the droplet is given by
\begin{equation}
  h(R_0)\approx -\frac{d'}{2}+\Big(\frac{d'^2}{4}+\frac{\Delta \sigma}{2\sigma_{\rm{sm}}}R_0^2\Big)^{1/2}
\label{eq:h}
\end{equation}

\noindent where $R_0$ is the droplet radius (Figure~\ref{fig:figure1}a),  $d'$ is the thickness of the surrounding film reduced by the thickness of the smectic layer covering the droplets, $\Delta\sigma = \sigma_{\rm{iso}} - \sigma_{\rm{sm}}$, $\sigma_{\rm{iso}}$ is the net interfacial tension of the isotropic droplet with respect to the gas phase, $\sigma_{\rm{sm}}$ is the smectic-gas interfacial tension. The equation~\ref{eq:h} disregards the exchange of the smectic material and the anisotropy of the surface tension in the smectic. A droplet remains stable for $\sigma_{\rm{iso}}>\sigma_{\rm{sm}}$. In the opposite case, one would expect a complete spreading of the droplet over the film area, accompanied by thinning and finally, rupture. A stable droplet shape is a pair of symmetric spherical caps (Figure~\ref{fig:figure1}a,b), which agrees well with the prediction of equation~\ref{eq:h}.  The surface of the droplet forms an apparent contact angle with the film surface, which weakly depends on temperature through the dependence of $\Delta \sigma(T)$. At a given temperature, the ratio $\epsilon=H_0/R_0$ of the height $H_0$ to the radius $R_0$  remains constant for any droplet (Figure~\ref{fig:figure1}c). This gives us natural lengthscales $R_0$ and $H_0$ to describe the changes of the droplet shape in horizontal and vertical dimensions.

\begin{figure}[htp!]
	\centering
	\includegraphics[width=0.7\linewidth]{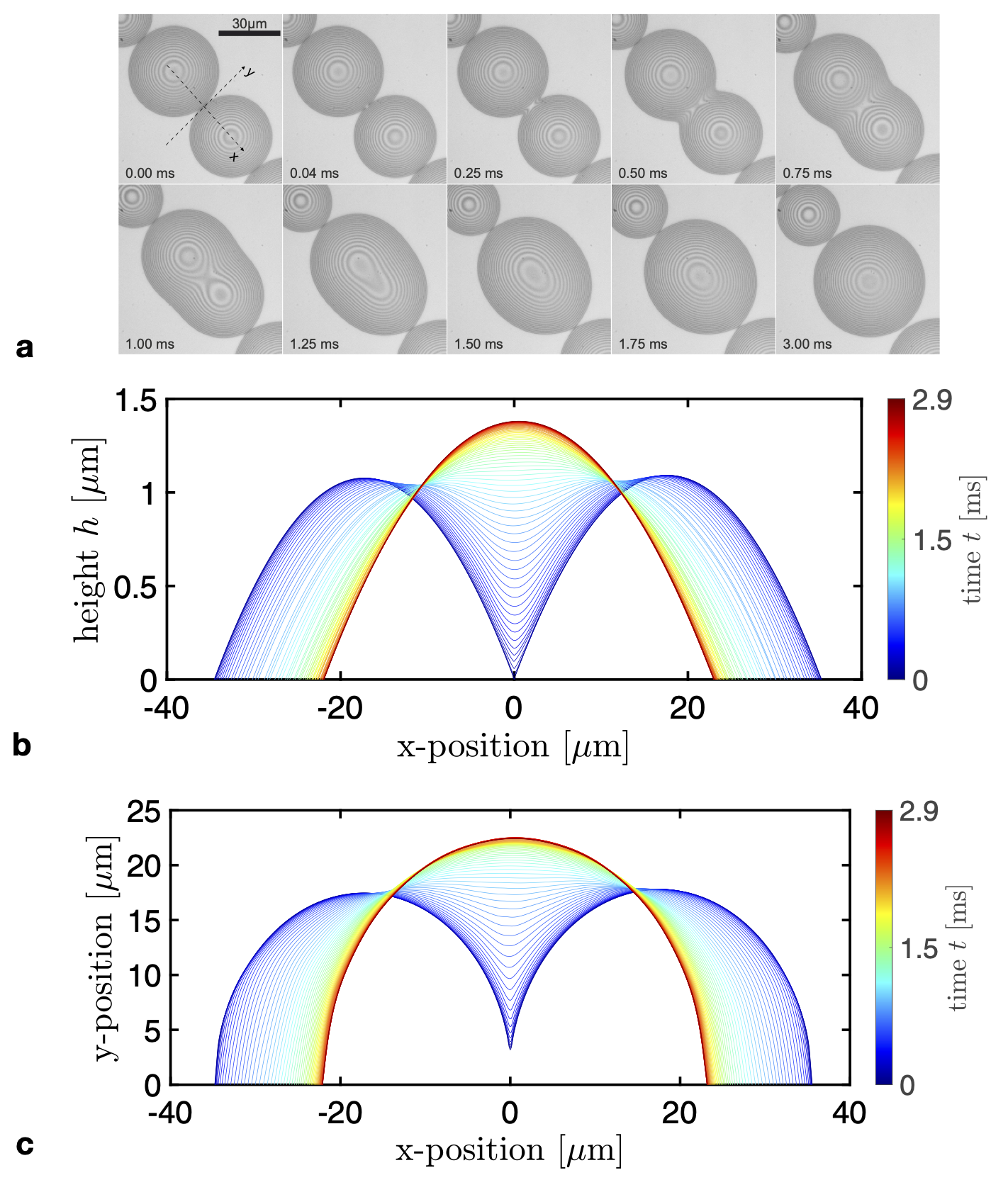}
	\caption{(a) A time sequence of images demonstrating merging droplets. (b) Vertical cross section of the droplets from (a) showing the displacement of the droplet boundaries and the growth of the bridge. (c) The horizontal cross section in the film plane (only one symmetric half $y>0$ is shown).} 
	\label{fig:profiles}
\end{figure}

After droplet nucleation, three scenarios have been observed: long-range repulsion, where the droplets form a regular lattice on the film surface~\cite{Klopp:2019gb}, short-range repulsion (Figure~\ref{fig:figure1}b), where the droplets remain in apparent contact, and the coalescence or merging (Figure~\ref{fig:profiles}a). 

The short-range repulsive state has a long life-time. The droplets can remain in close contact for hours. This repulsion can be attributed to the repulsion of the layer dislocations forming around the droplet boundaries. Activation energy is required to overcome this repulsive barrier $E_\mathrm{a} \approx 2000 k_\mathrm{B}T$. The coalescence can occur spontaneously or upon applying a transversal force to the droplets. This is achieved by a slight curving the film tuned by lowering the pressure in the film chamber (see Methods Section). 

\begin{figure}[h!]
	\centering
	\includegraphics[width=0.9\linewidth]{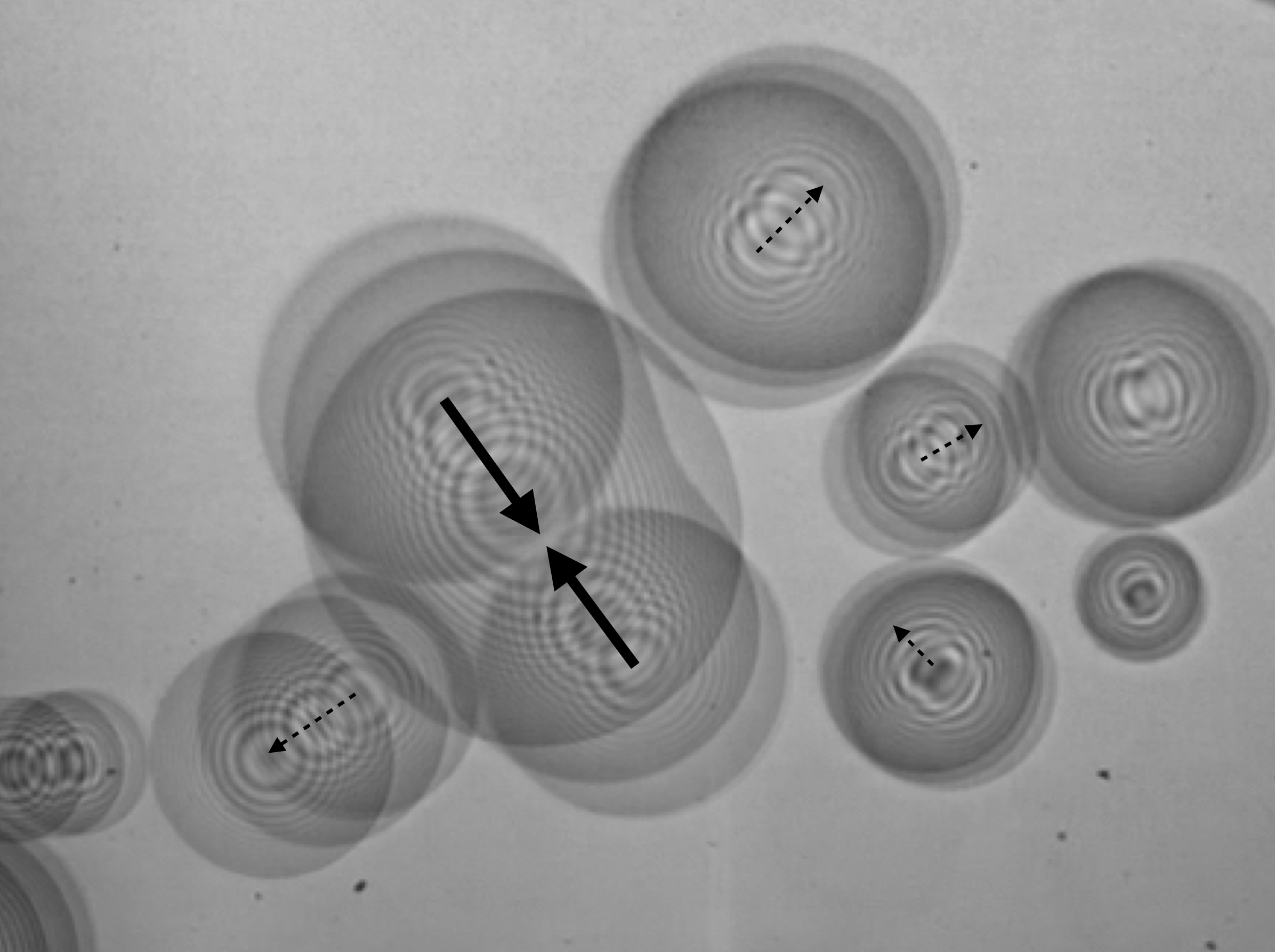}
	\caption{Superposition of three frames taken with an interval of 0.17 ms during the coalescence of a droplets pair. The solid arrows show the displacement of merging drops and the dashed arrow show the flow-induced displacements of the neighbouring droplets.
	} 
	\label{fig:flow}
\end{figure}

During the coalescence, a bridge forms between the pair of droplets and grows in width $W_\mathrm{b}(t)$ and height $H_\mathrm{b}(t)$ (Figure~\ref{fig:profiles}). The net volume of the droplets remains constant during the coalescence (time scale of a few milliseconds), provided the temperature is constant. As seen in figures~\ref{fig:profiles}b,c, the outer boundary of the merging droplets displaces inwards, so, that the contact angle remains nearly constant. In the horizontal cross section (Figure~\ref{fig:profiles}c), there is a pair of stationary points which do not experience any displacement. A similar situation was found in coalescence of 2D smectic islands \cite{Shuravin:2019eu}. The coalescence is accompanied by flow in the surrounding film as seen in figure~\ref{fig:flow}. The flow induced in the film during merging is able to displace and even deform the neighbouring droplets (see Supporting Information).

\begin{figure}[h!]
	\centering
	\includegraphics[width=0.9\linewidth]{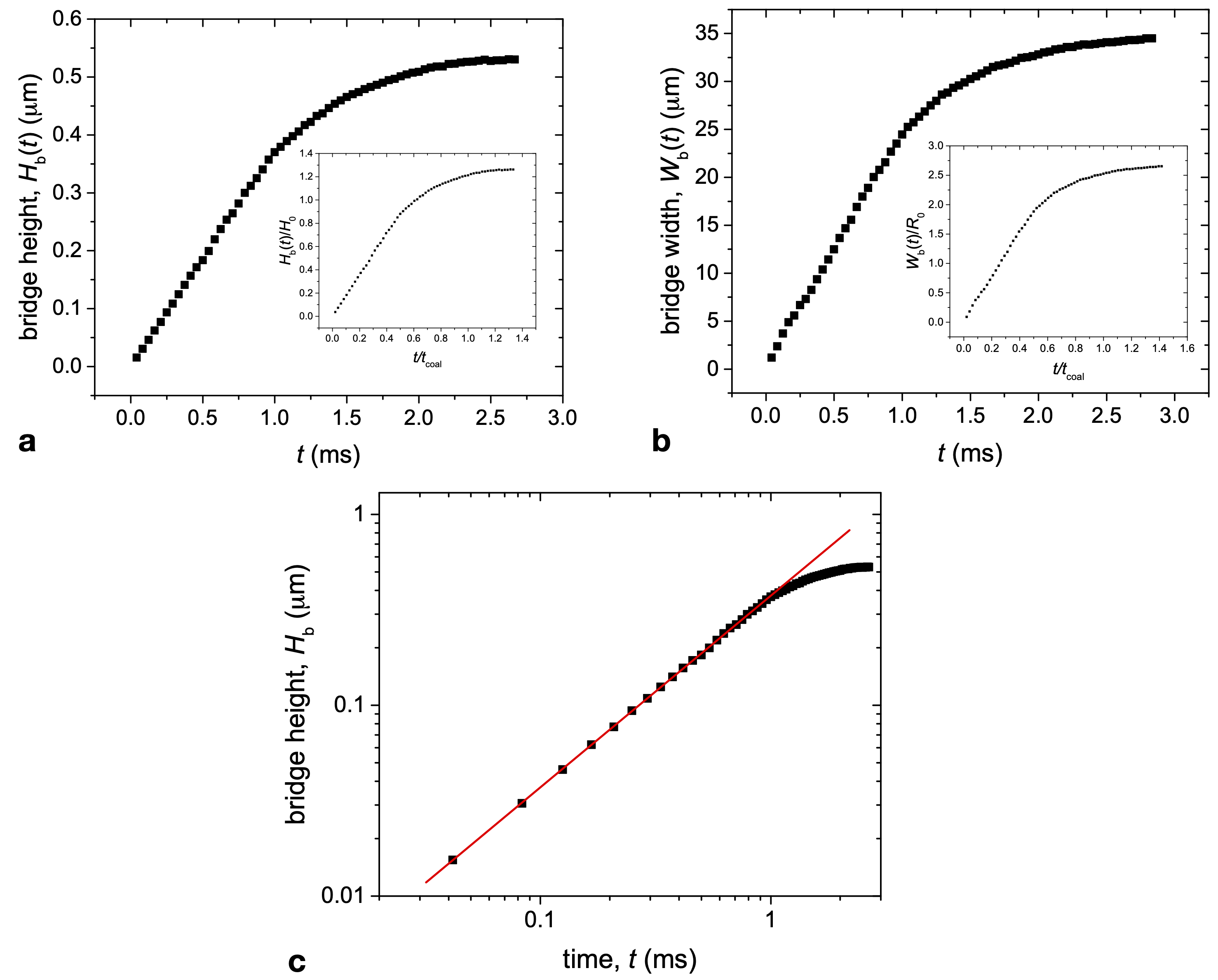}
	\caption{(a) Bridge height $H_\mathrm{b}(t)$ depending on time $t$ for a pair of droplets with the diameters and heights $D_{01}=35.5$ \textmu m, $H_{01}=1.13$ \textmu m and $D_{02}=34.7$ \textmu m, $H_{02}=1.10$ \textmu m. The inset shows the scaled bridge height  $h(t)=H_\mathrm{b}(t)/H_0$ depending on the scaled coalescing time. (b) Bridge width $W_\mathrm{b}(t)$ depending on time $t$. The inset shows the scaled bridge width $w(t)=W_\mathrm{b}(t)/R_0$ depending on the scaled coalescing time. (c) Bridge height in the double logarithmic scale. The red line is a linear fit with the slope $\alpha=1.0$.
	} 
	\label{fig:bridge}
\end{figure}

\begin{figure}[htp!]
	\centering
	\includegraphics[width=0.9\columnwidth]{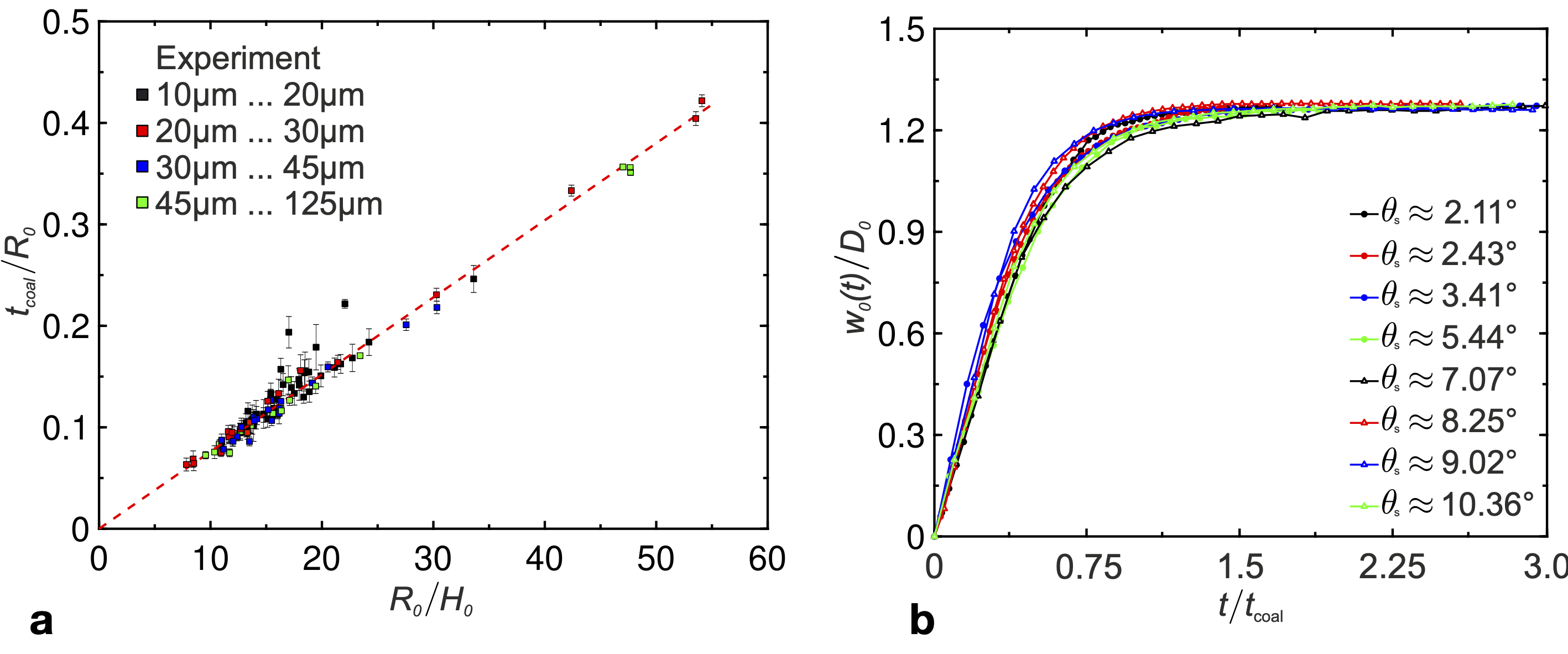}
	\caption{(a) Coalescence time $t_{\rm{coal}}$ normalised by the initial droplet radius $R_0$ as a function of $R_0/H_0$ for pairs of equal droplets. (b) Scaled bridge width as a function of scaled time for the coalescence events observed at different temperatures and different contact angles (see the text).} 
	\label{fig:coal_time}
\end{figure}

The coalescence rate can be determined by examining the time dependences of the bridge width $W_\mathrm{b}(t)$ and height $H_\mathrm{b}(t)$ depicted in the insets in figure~\ref{fig:bridge}.
The curves exhibit a linear growth $h(t)\propto t^{\alpha}$, $\alpha\approx 1$ at the initial stage, which turns to a saturating trend at longer times (Figure~\ref{fig:bridge}c). This suggests that the coalescence dynamics have an overdamped character. From this plot, we can estimate the coalescence time needed for a pair of droplets to merge (Figure~\ref{fig:coal_time}). 
 
Taking the time needed for the height $h(t)$ of the bridge to reach 95\% of the saturation as a criterium, we plot $t_{\rm{coal}}$ as a function of $\epsilon^{-1}=R_0/H_0$, where $H_0$ and $R_0$ are the initial height and the radius of the droplets  (Figure~\ref{fig:coal_time}a).  As seen from the plot, the data points condense on a straight line suggesting a universal relation between the coalescence time and the size of the droplets for all measured values of the contact angle and, hence, for different temperatures. It is the coalescence time that defines the timescale existing in the system. This timescale is reciprocal to the aspect ratio of the droplets and proportional to the radius: $t_\mathrm{coal}\propto R_0\epsilon^{-1}$, where the proportionality factor is reciprocal to the characteristic (capillary) velocity. Rescaling the time dependence of the bridge width $w(t)$ with $\tau$ from equation yields a master curve  shown in figure~\ref{fig:coal_time}b. The proportionality factor is determined by the material parameters such as the surface tension and the viscosity (capillary velocity). The contact angle and the aspect ratio $\epsilon$ are determined by the surface tension difference $\Delta \sigma$ (see equation~\ref{eq:h}). An essential question is to find what determines the timescales for the long-time coalescence behaviour.


\section*{Discussion}
The coalescence is primarily driven by the gain of the interfacial energy when two droplets merge. The dissipation occurs mainly due to the viscous flow in the droplets. The full description of the coalescence dynamics can be found by minimising the surface energy and solving the  Navier-Stokes equation for the hydrodynamic flow. The surface energy contributes to the pressure inside the droplets and an additional line tension at the droplet-film contact line. The full 3D model is computationally challenging, especially because of the ill-posed nature of the boundary condition at the contact line and the singularity at the initial stage of the bridge formation \cite{{Khodabocus:2018dw},Aarts:2005kt}. The problem can be significantly simplified if we adopt the lubrication approximation and reduce the problem to two dimensions by averaging the flow profile across the thickness of the film and neglecting the inertia. Such approach has been used to describe the coalescence of sessile droplets on a solid substrate \cite{Khodabocus:2018dw,Sellier:2009kz,Gaskell:2004ff}. In this model, the contact line singularity is relieved by the introduction of a precursor film stabilised by the disjoining pressure. In the case of our system, the precursor film has the same physical meaning as the freely suspended (FS) film itself.  To apply the lubrication model to the droplets in FS films, the model should be modified as discussed below. At the same, it is interesting to compare the two cases of sessile droplets and the film-embedded ones.

Assuming a small curvature of the droplet surface, the Laplace pressure can be approximated by $P_{\rm{L}}(x,y)=-\sigma\nabla^2 H$. The net pressure is expressed by the sum of the Laplace pressure $P_{\rm{L}}(x,y)$ and the disjoining pressure $\Pi(X,Y)$, given by \cite{Starov:1994fv,{Gaskell:2004ff}}
\begin{equation}\label{eq:D_Pressure}
  \Pi(X,Y)=\frac{(n-1)(m-1)}{d'(n-m)}\sigma(1-\cos (\theta_s))\Bigg[ \Bigg( \frac{d'}{H(X,Y)} \Bigg)^n -  \Bigg( \frac{d'}{H(X,Y)} \Bigg)^m\Bigg]
\end{equation}
\begin{equation}\label{eq:Net_Pressure}
  P=-\sigma \nabla^2 H(X,Y) -\Pi(X,Y)
\end{equation}
where $n=9$ and $m=3$ are the parameters of the interaction potential, $d$ is the film thickness, and $\theta_{\rm{s}}$ is the contact angle. This expression in a form of the Lenard-Jones potential stabilises the droplet shape with a predefined contact angle. 
The governing equation for the droplet height in lubrication approximation can be obtained from the mass conservation:
\begin{equation} \label{eq:mass}
  \frac{\partial H}{\partial t}=-\nabla\cdot Q,
\end{equation}

\noindent where $Q=\int_{d}^{d+H}(U,V)^T \mathrm{d} z$ and $U(X,Y,Z)$, $V(X,Y,Z)$ are the in-plane components of the flow velocity.
Natural length scale is given by the droplet geometry. The coordinates can be scaled by the initial radius $R_0$ and the initial height of the droplets $H_0$: $x=X/R_0$, $y=Y/R_0$, $z=Z/H_0$. The scaled droplet height and the film thickness are $h=H/H_0$ and $d=d'/H_0$, respectively, and the pressure is scaled by $P_0=\sigma H_0/R^2_0$.
After rescaling, we obtain for the reduced pressure $p(x,y)$:
\begin{equation}\label{eq:Scaled_P}
 p=\nabla^2 h -B\Bigg[\Bigg(\frac{d}{h}\Bigg)^n - \Bigg(\frac{d}{h}\Bigg)^m\Bigg],
\end{equation}
 where $B=2(n-1)(m-1)/d(n-m)$.

The coalescence of sessile droplets was extensively studied in \cite{Khodabocus:2018dw,Sellier:2009kz,Gaskell:2004ff,Eggers:1999dz}. Due to the non-slip boundary conditions at the bottom interface of the droplets and the free slip at the top interface, the governing equation for $h$ in lubrication approximation is~\cite{Gaskell:2004ff} 
\begin{equation}\label{eq:Scaled_H}
 \frac{\partial h}{\partial t} = \nabla \cdot (h^3 \nabla p),
\end{equation}
The characteristic coalescence time for sessile droplets is given by $\tau=\frac{3\eta}{\sigma} \frac{R_0^4}{H_0^3}=\frac{3\eta}{\sigma \epsilon^3} R_0$. For some typical values $R_0=50$ \textmu m, $\epsilon=0.2$, we obtain $\tau_{\rm{c}}$ in the range of 1 - 10 s. Droplets in freely suspended films, however, show a quite different behaviour. The characteristic coalescence time is in the range of milliseconds. Additionally, the dependence of the coalescence time $t_{\rm{coal}}$ on the aspect ratio $\epsilon$ is qualitatively different as shown in 
figure~\ref{fig:coal_time}a.

 The two cases drastically differ in the distributions of the flow velocities across the droplet thickness. Indeed, the isotropic droplets in freely suspended films can float freely and the non-slip condition in the mid plane is not applicable here. This suggests, that the transversal flow in the droplet rather than the vertical flow profile determines the coalescence dynamics. The boundary conditions on both sides of the droplet surface becomes drag free (for the normal derivatives at the boundaries, $\partial u/\partial n = \partial v/\partial n=0$) and $\partial u/\partial z=\partial v/\partial z=0$. This results in a nearly uniform flow profile $u(x,y)$ and $v(x,y)$ satisfying 
\begin{equation}
\label{eq:px}
  \frac{\partial^2 u}{\partial x^2} + \frac{\partial^2 u}{\partial y^2}=\frac{\partial p}{\partial x}
\end{equation}
\begin{equation}
\label{eq:py}
  \frac{\partial^2 v}{\partial x^2} + \frac{\partial^2 v}{\partial y^2}=\frac{\partial p}{\partial y}
\end{equation}

\noindent when the inertial terms in the Navier-Stokes equation are neglected. Integration of $Q$ in the Eqn. \ref{eq:mass} yields 
\begin{equation}   
\label{eq:lubnew}
  \frac{\partial h}{\partial t}=  \Big(\frac{\partial hu}{\partial x} + \frac{\partial hv}{\partial y} \Big)
\end{equation}

\noindent where the characteristic time scale is determined by the capillary velocity $c$ and given by 
\begin{equation}   \label{eq:time2}
  \tau=\frac{\eta}{\sigma}\frac{R_0}{\epsilon}=c^{-1}\frac{R_0}{\epsilon}
\end{equation}
This new time scale exhibits exactly the same dependence on the radius $R_0$ of the droplet and the aspect ratio $\epsilon$ as observed in the experiments figure~\ref{fig:coal_time}. Taking the typical values for the material parameters $\eta$, $\sigma$ and droplet size $R_0$ and $\epsilon$, we obtain the coalescence time in the range of few milliseconds.  

\begin{figure}[htp!]
	\centering
	\includegraphics[width=0.9\columnwidth]{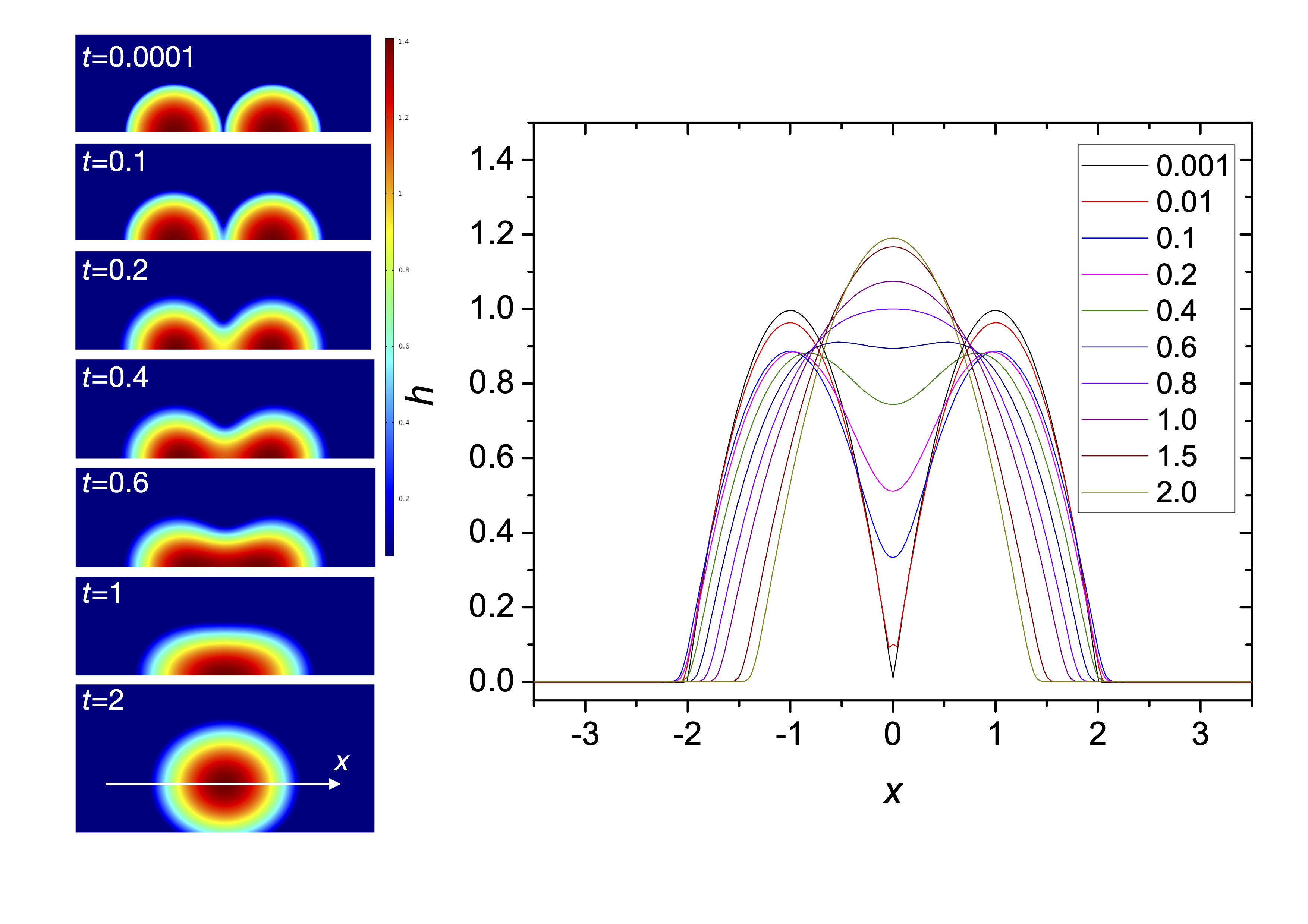}
	\caption{(Left) Simulated profiles of the merging droplets for different values of the dimensionless time $t$. (Right) Droplet cross-sections along the line connecting their centres. The legend shows the value of the dimensionless time $t$.} 
	\label{fig:mod_profiles}
\end{figure}

Equation~\ref{eq:lubnew} together with \ref{eq:D_Pressure},\ref{eq:px} and \ref{eq:py} describes the coalescence dynamics of isotropic droplets in freely suspended films in the lubrication approximation. The equations were solved using the finite element technique in the commercial software package COMSOL. As initial condition, we choose a pair of droplets of radii and heights equal to unity. The simulation results are shown in figure~\ref{fig:mod_profiles}. 
  
The simulation reproduces the major features observed in the experiment. The formation of the bridge is accompanied by the displacement of the droplets which takes place in the range of the dimensionless time $t\approx 1.5$. The net volume remains conserved in accordance with equation~\ref{eq:lubnew}. Comparing it with the experimentally measured coalescence time in the range of few milliseconds, we can estimate the capillary velocity $c\approx 2.4$ m/s.

 \begin{figure}[htp!]
	\centering
	\includegraphics[width=0.6\columnwidth]{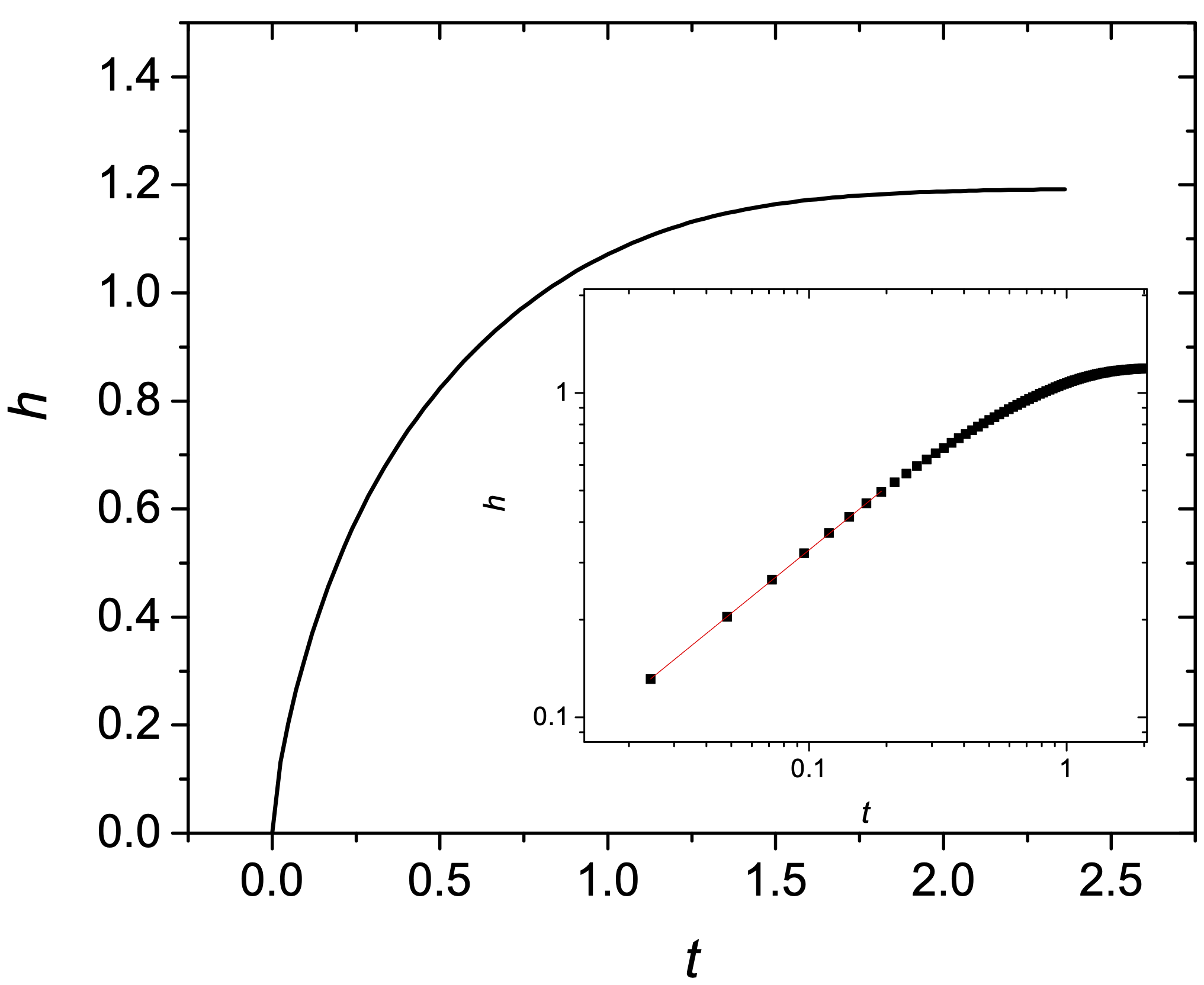}
	\caption{Simulated dimensionless bridge height $h$ as a function of the dimensionless time $t$. The inset is the log-log presentation of the plot.} 
	\label{fig:mod_h}
\end{figure}

The dependence of the dimensionless bridge height $h(t)$ is given in figure~\ref{fig:mod_h}. $h(t)$ is strongly nonlinear in the whole time range of the coalescence. In the intermediate time range, $h(t)\propto t^\alpha$ with $\alpha \approx 0.65$. Despite the fact that the model describes well the qualitative dependence of the coalescence time on the size and shape of the droplets, it greatly overestimates the coalescence rate in the beginning. The capillary velocity can be estimated comparing the experimental coalescence time $t_\mathrm{coal}$ with the model prediction. This yields $c=0.4$ m/s. The exact value of the capillary velocity is not known in the overheated isotropic droplets. However, one could estimate it using the material parameters of the liquid crystal. The surface tension measured in this compound $\sigma = 2.4\times 10^{-2}$ N/m and the viscosity $\eta=0.014$~Pa$\cdot$s, which yields $c=1.7$ m/s. This value is much higher than that assumed in the model suggesting that either $\sigma$ is too large or $\eta$ is too small. The stability condition against complete wetting requires that effective interfacial tension of the droplet interface is larger than that of the film. On the other hand, the viscosity $\eta$ changes only by a few percents upon the transition from the smectic to the isotropic phase. A rough estimation of the expected viscosity from the model capillary velocity and the smectic surface tension $\sigma = 2.4\times 10^{-2}$ N/m, yields $\eta \approx 0.06$~Pa$\cdot$s. This exceeds the smectic value by a factor of four. Additional contributions to the model are required to better describe the coalescence dynamics. One possibility to explain the discrepancy is to account for additional dissipation due to the reordering of the smectic layers. The coalescence is accompanied by a decrease of the droplet area. This is necessarily accompanied by the retraction of the smectic material to the film or exchange with the isotropic reservoir.

%
%

In summary, we experimentally investigated the coalescence of isotropic droplets in overheated smectic films. We demonstrated that the dependence of the coalescence time on the initial radius $R_0$ and the aspect ratio $\epsilon$ of the droplet suggests that the transversal velocity is nearly independent of the vertical coordinate. The coalescence time scales as 
$t_{\mathrm{coal}} \propto R_0/ \epsilon$. This result can be explained using lubrication approximation as a consequence of the independence of the flow velocity field on the vertical coordinate. However, a simple 2D model based on the lubrication theory fails to describe the behaviour of the initial coalescence regime and underestimates the coalescence time. Additional studies are required to understand the discrepancy of the theoretical capillary velocity.

\section*{Methods}

\subsection{Method subsection.}

\begin{figure}[h!]
	\centering
	\includegraphics[width=0.9\linewidth]{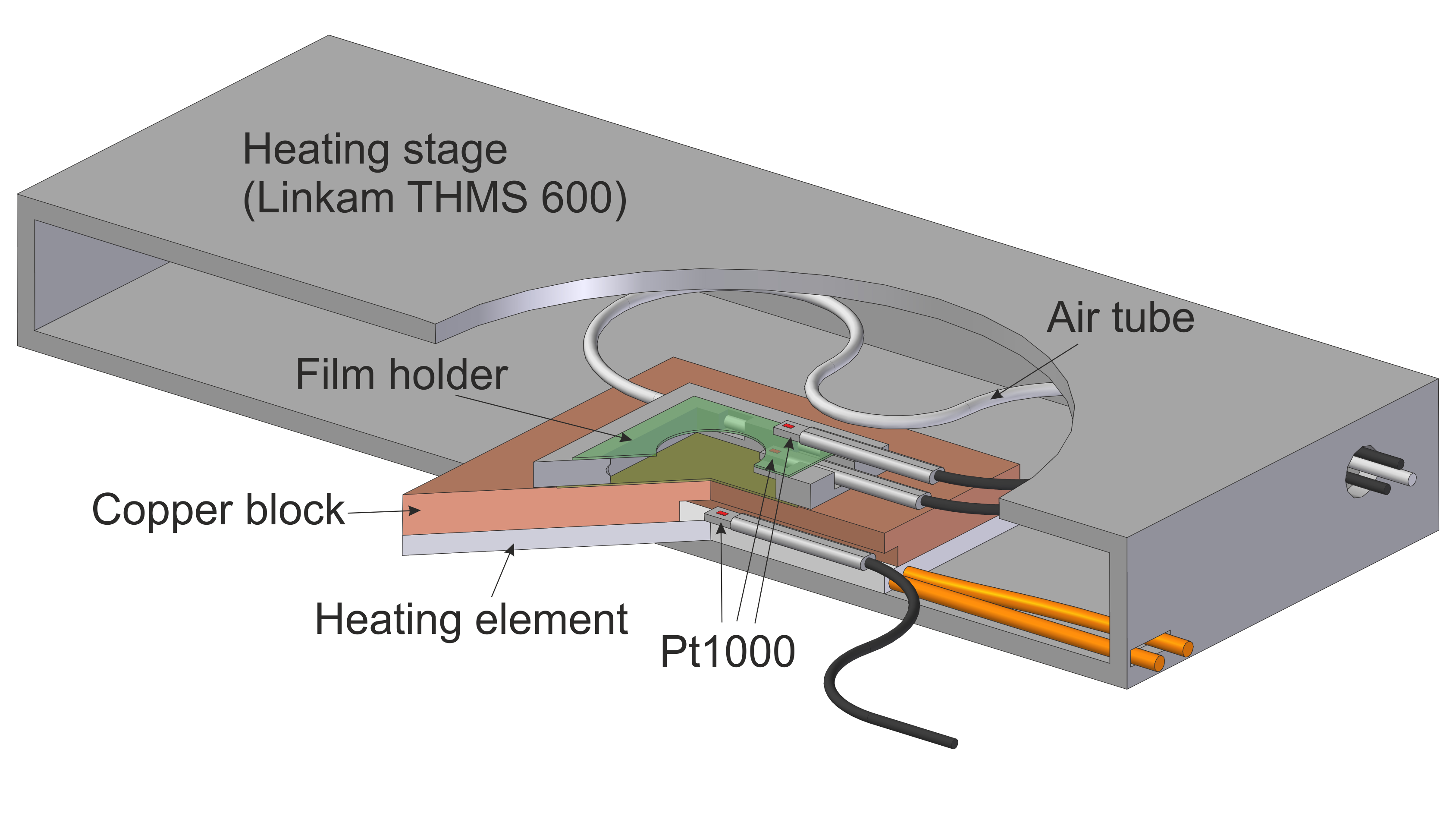}
	\caption{Film chamber for the preparation of freely suspended films. The body dimensions of the chamber are 160 $\times$ 80 $\times$ 24 mm. The diameter of the opening for the smectic film is 1 mm. }
	\label{fig:setup}
\end{figure}

For our experiments, we used the liquid crystal material mixture (MX 12160) containing 5-Heptyl-2-[4-(4-methylhexyloxy)-phenyl]-pyrimidine (80$\%$) and 4-(5-Octyl-pyrimidin-2-yl)-benzoic acid decyl ester (20$\%$). This mixture has a direct phase transition from the smectic-A (SmA) phase at room temperature to the isotropic phase at 54$^\circ$ C. During our experiments, we used the film chamber shown in figure \ref{fig:setup} with a film holder placed in a closed heating stage to avoid air flow from the outside. The curvature of the film was adjusted by tuning the pressure $p$ below the film ($-50$ Pa $\leq p \leq$ $50$ Pa).
By that, we created an effective gravitational force acting on the droplets in the regions where the film was inclined to the horizontal. The temperature was measured by three PT100 sensors. 
 We used a HALJIA heating element with an extension of 40mm x 40mm and the maximum power of 48W at 12V.\\

The film chamber was placed under a polarising microscope (ZEISS Axioscope 40), where we added a mercury lamp, a 546 nm narroband filter and the Phantom VEO 710L high-speed camera to observe the coalescence with a frame rate of 24000 fps with a typical frame size of 500 x 500 pixels.
The film thickness $h$ (in the range from $50$ nm  up to $1500$ nm) was determined from the reflectivity spectrum $R(\lambda)$ measurements. The thickness profile of the droplets was determined using interferometry in monochromatic light.



The numerical simulations were made using COMSOL finite element package. The equations were solved in the weak form. The expression~(\ref{eq:lubnew}) multiplied by a test function $\psi(x,y)$ and integrated over the whole rectangular domain $\Omega$. The test function was chosen to satisfy the condition at the domain boundary $\psi(x,y)|_{\partial \Omega}=0$. This yields 
\begin{equation}
  \int_{\Omega} \Big(\psi\frac{\partial h}{\partial t} -\frac{\partial \psi}{\partial x} h u  -\frac{\partial \psi}{\partial x} h v\Big)\mathrm{d}\Omega =0
\end{equation}
The equations \ref{eq:D_Pressure},\ref{eq:px} and \ref{eq:py} were converted in the weak form in the similar way.





\bibliographystyle{unsrtnat}


\begin{addendum}
 \item The authors thank Professor Ralf Stannarius for fruitful discussions and helpful comments.  This research was supported was supported by the German Aerospace Center (DLR) with project 50WM1744 and by the Deutsche Forschungsgemeinschaft (DFG) with project STA 425/40-1. C. K. acknowledges support by a Landesstipendium Sachsen-Anhalt and DAAD Program PPP USA. 
 
 \item[Competing Interests] The authors declare that they have no
competing financial interests.
 \item[Correspondence] Correspondence and requests for materials
should be addressed to A.E.~(email: alexey.eremin@ovgu.de).
\end{addendum}


\end{document}